\documentclass[fleqn,twoside]{article}
\usepackage{espcrc2}
\usepackage{graphicx}
\title{Renormalization of an effective model Hamiltonian by a counter term}
\author{Michael Frewer\address[1]{%
        Max-Planck-Institut f\"ur
        Kernphysik, D-69029 Heidelberg, Germany}, 
        Tobias Frederico\address{%
        Dep. de F\'\i sica, Instituto Tecnol\'ogico de Aeron\'autica, 
        Centro T\'ecnico Aeroespacial,\\ 
        12.228-900, S. Jos\'e dos Campos, SP, Brazil} 
  and 
        Hans-Christian Pauli\addressmark[1]} 
\begin{document}
\begin{abstract}
An ill-defined integral equation for
modeling the mass-spectrum of mesons is regulated with  
an additional but unphysical parameter.
This parameter dependance is removed by renormalization.
Illustrative graphical examples are given.
\vspace{1pc}
\end{abstract}
\maketitle

We focus on the integral equation
\begin{eqnarray*}
 \begin{array}{@{}l@{}c@{}l@{}} 
   \Big[M^2 - 4m^2 - 4\vec k ^2 \Big]\phi(\vec k)
   &=& \displaystyle 
   \int \!\! d^3\vec k'\ U (\vec k',\vec k)
   \phi(\vec k') ,
 \end{array}\!\!\!\!\!\!\!\!\!\!
\end{eqnarray*}
\noindent
with the attractive kernel
\begin{eqnarray*}
U(\vec k',\vec k) = -\frac{4}{3\pi^2} \frac{\alpha}{m}
    \left[\frac{2m^2}{(\vec k' -\vec k)^2} 
+ 1\right]
.\end{eqnarray*} 
It has two parameters $\alpha$, $m$. 

From a physical point of view the equation is a QCD-inspired 
effective one-body bound-state equation for
modeling mesons with different constituent quark flavors \cite{Pau00}.
$M^2$ are the invariant mass squares of the physical mesons, while
$m=m_1=m_2$ is the effective mass of the quark and anti-quark.
It takes this explicit
form due to an over-simplification by the  
$\uparrow\downarrow$-model \cite{Pau00}. 
If one Fourier tranforms the  
kernel $U$ to configuration
space, the interaction potential
consists of a long-ranged Coulomb-interaction
and a short-ranged delta-interaction.
It is this
latter part, which generates all the well known trouble. 
In order to get reasonable solutions one has to
regulate the high momentum transfers $Q^2=(\vec k' -\vec k)^2$.
Therefore we substitute the number 1 
by a regulating function, $1\rightarrow R(\Lambda,Q)$, 
for which the soft cut-off
\begin{eqnarray*}
   R(\Lambda,Q^2)=\frac{\Lambda^2}{\Lambda^2+Q^2}
=\frac{\Lambda^2}{\Lambda^2+(\vec k'-\vec k)^2}
\end{eqnarray*}
is chosen.
In configuration
space the short-ranged delta is now
smeared out to a Yukawa interaction. 
Since the regulator $\Lambda$ is an additional
but unphysical parameter, one has to renormalize 
the equation in order to restore the original problem
in the limit $\Lambda\rightarrow\infty$. 
For getting a greater tranparency we want to interpret 
the physical parameters
$\alpha$ and $m$ as renormalization constants.

That we are dealing here with a bound-state equation
on the light-cone, can not be seen explicitely. 
The above equation results from a variable transform in the 
longitudinal momentum fraction $x$.
For equal masses the relation is given by \cite{Pau00}
\begin{eqnarray*}
   x(k_z)=\frac{1}{2}\left(
   1+\frac{k_z}{\sqrt{m^2+\vec k^2_\perp +k_z^2}}\right)
.\end{eqnarray*}
The relationship between the light-cone wavefunction $\psi$
and the function $\phi$ is given by
\begin{eqnarray*}
   \psi(x,\vec k _{\!\perp}) =
   \frac{\phi(\vec k)}{\sqrt{x(1-x)}}
   \left[1+\frac{\vec k ^2}{m^2}\right]^\frac{1}{4}
.\end{eqnarray*}
The function $\phi$ has no physical meaning in the sense of a
probalility amplitude and is refered to as the
reduced wavefunction.

\begin{figure}[t]
\begin{center}\scalebox{0.40}{
\includegraphics{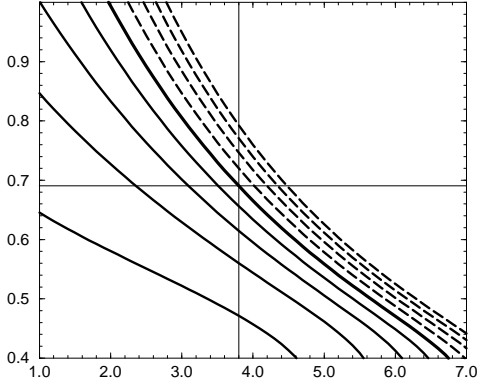}
}\end{center} 
\vspace{-0.75cm}
\caption{\label{fig:bild1} 
   Nine contours $0.4\leq\alpha_n(\Lambda)\leq 1.0$ are 
   plotted versus $1.0\leq\Lambda/\Delta \leq 7.0$ 
   from bottom to top with $n=4,3,\cdots ,-3,-4$.
   The contours are obtained by 
   $M_0^2(\alpha,\Lambda) = n\Delta^2 + M_{\pi}^2$. 
   The thick contour $n=0$ describes the pion with 
   $M_0^2=M_{\pi}^2$.
   Masses are given in units of $\Delta=350$~MeV.  
}\end{figure}

\begin{figure}[t]
\begin{center}\scalebox{0.40}{
      \includegraphics{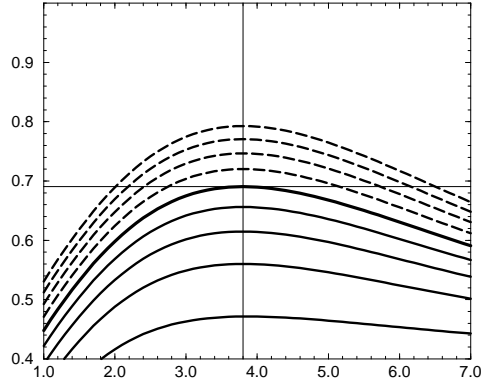}
}\end{center} 
\vspace{-0.75cm}
\caption{\label{fig:bild2} 
    Nine contours $0.4\leq\alpha_n(\Lambda)\leq 1.0$ are plotted
    versus $1.0\leq \Lambda/\Delta\leq 7.0$ from bottom to top 
    with $n=4,3,\cdots ,-3,-4$. 
    Here the partially renormalized 
    $M_0 ^2(\Lambda,\alpha)=n\Delta^2 + M_\pi^2$ with 
    $\Delta=350$~MeV is displayed by contours.
    The thick contour $n=0$ describes the pion with 
    $M_0^2=M_\pi^2$.}
\end{figure}

After regularization one faces an integral equation with three
parameters $\alpha$, $m$ and $\Lambda$.
For simplification the functions $\phi$ are restricted to
the calculation of s-waves: $\phi(\vec k)=\phi(|\vec k|)$
and by reasons explained below, we fix $m=406$~MeV.

The spectrum of the bound-state mass squares 
$M_i^2(\alpha,\Lambda)$ are then calculated numerically.
For the ground state $M_0^2(\alpha,\Lambda)$ this is displayed
in Figure 1. A similar graph could be given for the first 
excited state $M_1^2(\alpha,\Lambda)$. 

\section{Example for local renormalization}
The new parameter $\Lambda$ appears due to
regularization. According to renormalization
theory the spectrum may not depend on this formal
parameter, thus we must require
\begin{eqnarray}
   \delta_\Lambda M^2(\Lambda) \stackrel{!}{=} 0.
\end{eqnarray}
To achieve this, we extend the model interaction by a 
adding to $R$ a counter term $C(\Lambda,Q)$.
We choose this function according to three criteria.
First, the new function $\widetilde R\equiv R + C$ 
must again be a regulator.
Second, we require that a zero is added for a particular 
value of $\Lambda$, say for $\Lambda=\mu$.
Third, we require the first $\Lambda$-derivative of
$\widetilde R$ to vanish at $\Lambda=\mu$.
The conditions are met by 
\begin{eqnarray*} 
    C(\Lambda,Q)  = -Q^2\frac{(\Lambda^2-\mu^2)}{(\Lambda^2+Q^2)^2}
.\end{eqnarray*}
The kernel of the integral equation becomes then
\begin{eqnarray*} 
    U &=& -\frac{4}{3\pi^2}         
    \frac{\alpha}{ m}
    \left(\frac{2m^2}{Q^2} + R + C \right) 
.\end{eqnarray*}
The lowest eigenvalue of the corresponding integral equation 
is displayed in Figure 2 as function of $\alpha$ and $\Lambda$.

Based on the Hellmann-Feynman theorem, and 
\begin{eqnarray*}
\frac{d\widetilde R}{d\Lambda^2} =
  2Q^2\frac{(\Lambda^2-\mu^2)}{(\Lambda^2+Q^2)^3}
,\end{eqnarray*}
one expects that the derivative of the eigenvalues
change sign at $\Lambda=\mu$.
The numerical results in Figure 2 illustrate this
very convincingly. 
In fact, for the numerical value $\mu=1330$~MeV $(\mu/\Delta=3.8)$,
the eigenvalues satisfies Eq.(1).
The Hamiltonian is thus partially renormalized
in the vicinity of $\Lambda \sim \mu$ for all $\alpha$.

\section{Exact renormalization by counter terms}
Above, we have constructed a local
renormalization counter term in the region of $\Lambda/\Delta=3.8$.
Now our aim is to renormalize globally, i.e. for
all possible $\Lambda$. This can be achieved
by requiring that the $\Lambda^2$-derivatives 
of all orders have to vanish in the point $\Lambda=\mu$. 
Besides that, we will take up an easier
and more straightforward way to derive a global counter term.

The regularization function $\widetilde R$ is defined by:
\begin{eqnarray*}
    \widetilde R(\Lambda,Q) =
    R(\Lambda,Q) + C(\Lambda,Q)
,\\ \quad\mathrm{with}\quad
    R(\Lambda,Q) = \frac{\Lambda^2}{\Lambda^2+Q^2}
.\end{eqnarray*} 
Goal is to construct a counter term $C$ such that
\begin{eqnarray*}
    C(\Lambda=\mu,Q)=0
,\quad\mathrm{and}\quad
    \left.\frac{d\widetilde R}{d\Lambda^2}\right|_{\forall \Lambda} = 0
.\end{eqnarray*} 
The requirements are satisfied by the
differential equation
\begin{eqnarray*}
    \frac{dC}{d\Lambda^2} = - \frac{dR}{d\Lambda^2} = -
     \frac{Q^2}{(\Lambda^2+Q^2)^2}
.\end{eqnarray*} 
The boundary conditions are included by its
integral form
\begin{eqnarray*}
    C(\Lambda,Q) & = & - \int\limits_{\mu^2}^{\Lambda^2}
    d\lambda^2 \frac{dR(\lambda^2,Q)}{d\lambda^2}
\\ & = &
      \frac{\mu^2}{\mu^2+Q^2} -
     \frac{\Lambda^2}{\Lambda^2+Q^2}
.\end{eqnarray*}
The regularization function
$\widetilde R$ becomes
\begin{eqnarray*}
    \widetilde R(\Lambda,Q) =
    \frac{\mu^2}{\mu^2+Q^2}
,\!\!\!\!\end{eqnarray*} 
which is to be used in the integral equation
of the $\uparrow\downarrow$-model, i.e.
\begin{eqnarray*}
 \begin{array}{@{}l@{}l@{}} 
   &\Big[M^2 - 4m^2 - 4\vec k ^2 \Big]\phi(\vec k)
   = \displaystyle 
   \int \!\! d^3\vec k'\ U (\vec k',\vec k)
   \phi(\vec k'),
\\ &\mathrm{with}\quad
   U (\Lambda,Q) = - \displaystyle 
   \frac{4}{3\pi^2} \frac{\alpha}{m}
   \Big(\frac{2m^2}{Q^2} +\frac{\mu^2}{\mu^2+Q^2}\Big).
 \end{array}
\hspace{-5em}
\end{eqnarray*} 
The equation is now manifestly independent of
$\Lambda$ and the limit $\Lambda\rightarrow\infty$
can be taken trivially.
In line with the theory of renormalization, the
three parameters $\alpha$, $\mu$ and $m$ 
have to be determined by experiment, i.e. in principal
three exprimental values are needed to fix them.

%
\begin{figure}[t]
\begin{center}\scalebox{0.40}{
\includegraphics{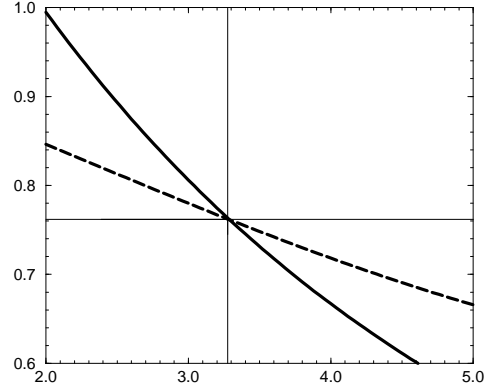}
}\end{center} 
\vspace{-0.75cm}
\caption{\label{fig:bild3} 
Two contours $\alpha_n(\mu)$ are plotted
versus $2.0\leq \mu/\Delta\leq 5.0$, with
$\Delta=350$~MeV.      
The solid contour $\alpha_\pi(\mu)$ is obtained by fixing
the lowest eigenvalue to the pion ground state
$M^2_0=M^2_{\pi}=(140\mbox{~MeV})^2$, while the dotted contour
$\alpha_{\pi^*}(\mu)$ refers to the fixing of the  
second lowest eigenvalue to the first exited state of the pion 
$M^2_1=M^2_{\pi^*}=(768\mbox{~MeV})^2$.
}\end{figure}
%

\section{Determining the parameters $\alpha$ and $\mu$}
We fix the two unknown parameters $\alpha$
and $\mu$ by the experimental values of the
ground and excited state mass of the pion.
The pion has the mass $M_\pi=140$~MeV. The
precise empirical value of the excited pion mass
is not known very well. We
choose here $M_{\pi^*}=M_\rho=768$~MeV
for no good reason other than convenience.
This large value is the reason for our
comparatively large quark mass $m=406$~MeV,
which is fixed here once and for all.

Each of the two equations, 
$M_0^2(\alpha,\mu)=M_\pi^2$ and 
$M_1^2(\alpha,\mu)=M_{\pi^*}^2$ 
determine a function $\alpha(\mu)$,
as illustrated in Figure 3.
Their intersection point determines the solution,
that is  
$\alpha_0=0.761$ and $\mu_0=1.15$~GeV,
or $\mu_0/\Delta \sim 3.28$, as displayed in the figure.

Important to note is that the two contours
in Figure 3 are intersecting only once.
This crossing of the contours is
unique, even for $\mu\rightarrow\infty$.

\textbf{Acknowledgment.}
Michael Frewer enjoys to thank the organizers
of the meeting for support.

\end{document}